\documentclass[12pt, prd, showpacs]{revtex4}
\usepackage{amssymb}
\usepackage{amsmath}

\input{tcilatex}

\begin{document}

\title{Decay of charged particles near naked singularities and super-Penrose
process without fine-tuning}
\author{O. B. Zaslavskii}
\affiliation{Department of Physics and Technology, Kharkov V.N. Karazin National
University, 4 Svoboda Square, Kharkov 61022, Ukraine}
\email{zaslav@ukr.net }

\begin{abstract}
We consider the Penrose process near the naked singularity in the
Reissner-Nordstr\"{o}m metric. Particle 0 falls from infinity and decays to
two fragments at some point $r_{0}$. We show that the energy extraction due
to this process can be indefinitely large in the limit $r_{0}\rightarrow 0$.
In doing so, the value of the particle charge can remain bounded, in
contrast to the previously known examples of the Penrose process in the
electric field with unbounded energy extraction. The effect persists even in
the limit of the flat space-time. Uhbounded energy gain is obtained in the
standard (not collisional) Penrose process and does not require fine-tuning
of particle parameters or characteristics of space-time.
\end{abstract}

\maketitle

\section{Introduction}

The Penrose process (PP) \cite{pen} is one of the remarkable physical
effects typical of general relativity and other theories of gravity. Let us
suppose that in a space-time there exists a region (called ergosphere) where
the Killing energy $E$ measured at infinity can be negative. If a particle 0
within the ergosphere splits to two fragments 1 and 2, one of them can have $%
E_{1}<0$. Then, the conservation of energy entails that the second fragment
has $E_{2}>E_{0}$, so some energy gain occurs. This is just the PP. However,
it has a quite modest efficiency in realistic astrophysics. Its efficiency
increases if instead of decay of one particle \ the process consists in
collision of different particles. Such a process with energy gain is called
the collisional Penrose process (see Ref. \cite{schrev} for review and
references therein).

Especially, it is relevant in the context of another effect that is called
the Ba\~{n}ados-Silk-West (BSW) one \cite{ban}. It happens if particles
collide near the black hole. Then, under certain conditions the energy $%
E_{c.m.}$ in the center-of-mass frame becomes unbounded. Meanwhile, the gain
in the energy $E$ remains bounded \cite{j} - \cite{z}.

Originally, all these effects were found in the vicinity of rotating black
holes. However, the similar effect should occur in the background of static
charged black holes \cite{ruf}. It was investigated further in detail \cite%
{den} - \cite{dad3}. Moreover, there exists a limit to flat space-time when
the effect under discussion persists \cite{df}, \cite{flat}. It amounts to
the possibility of indefinitely large energy gain when $E$ is formally
unbounded (in the test particle approximation , when backreaction on a
metric is neglected). Then, such a process is called the super-Penrose one
(SPP). In contrast to collisions of neutral particles near rotating black
holes where the SPP\ is forbidden (see, e.g. Ref. \cite{is} and references
therein), the SPP\ is indeed possible for the charged black holes, even for
pure radial particle motion in the Reissner-Nordstr\"{o}m (RN) background 
\cite{rn}, \cite{nem}. (More general scenarios can include both the electric
charge and rotation \cite{hlz}.)

In the present work, we show that there exists one more type of the SPP. It
is realized in the background of naked charged singularities. The SPP\ near
naked singularities was already the subject of study for rotating geometries
with neutral particles. In \cite{inf}, it was shown that the SPP is possible
for the Kerr overspinning metric when particles move within the equatorial
plane. In \cite{tzw}, this result was generalized to generic rotating
axially symmetric space-times with naked singularities. In \cite{n}, high
energy extraction in the Kerr overspinning geometry was investigated for
particle collisions taking place on the off-equatorial planes. These papers
share two common features: (i) the unbounded (or at least very high) energy
extraction requires fine-tuning between the parameters of geometry, so it is
on a threshold of forming a horizon (but the horizon does not form), (ii)
the effect under discussion implies the collisional Penrose process. 

We would like to stress that in the scenarios considered in the present
work, both conditions are relaxed: the effect is achieved for quite generic
geometries without special conditions for particles and, also, it happens
for the decay typical of the standard Penrose process instead of the
collisional one.

The process with charged particles in the RN background discussed in the
present work differs not only from its counterpart in the rotating backgound
with neutral particles but also from its analogs near the RN naked
singularity considered earlier \cite{naked}, \cite{ns}. In aforementioned
papers (i) collisions of shells were studied, (ii) the high energy process
implies high $E_{c.m.}$ Meanwhile, we show that (i) the effect under
discussion is valid for test particles, (ii) it involves ultra-high energies 
$E$, (iii) there are no collisions at all and the process represents a
standard PP, not a collisional one.

There is one more interesting property inherent to the scenarios considered
in the present paper. The examples of the electric SPP, known before, share
the common feature. If one wants to gain large energy, the electric charge
is also should be large. This concerns both the standard and collisional PP
as well as the confined one \cite{conf}. However, the electric charge of
elementary particles, atoms or nuclei cannot be arbitrary large \cite{rn}, 
\cite{axis}. Also, there exist similar restrictions for macroscopic bodies 
\cite{nem}. Meanwhile, we demonstrate that for naked singularities not only
PP but also SPP does exist for a finite value of a particle charge.

We use the geometric system of units in which fundamental constants $G=c=1$.

\section{Basic equations}

We consider the Reissner-Nordstr\"{o}m metric%
\begin{equation}
ds^{2}=-dt^{2}f+\frac{dr^{2}}{f}r^{2}d\omega ^{2}\text{,}
\end{equation}%
where $d\omega ^{2}=d\theta ^{2}+\sin ^{2}\theta d\phi ^{2}$,%
\begin{equation}
f=1-\frac{2M}{r}+\frac{Q^{2}}{r^{2}}\text{.}
\end{equation}%
Here $M$ and $Q$ are the mass and electric charge, respectively. We take $%
Q>0 $. We assume that $M<Q$, so there is a naked singularity at $r=0$ in
this space-time.

Now, we consider motion of test particles. We restrict ourselves by pure
radial motion. This is sufficient to demonstrate that the effect under
discussion does exist. Then, equations of motion read%
\begin{equation}
m\dot{t}=\frac{X}{f},
\end{equation}%
\begin{equation}
X=E-q\varphi \text{,}  \label{X}
\end{equation}%
\begin{equation}
m\dot{r}=\sigma P\text{, }P=m\sqrt{U}\text{, }U=\frac{X^{2}}{m^{2}}-f,
\label{P}
\end{equation}%
$\sigma =\pm 1$ depending on the direction of motion, dot denotes
differentiation with respect to the proper time $\tau $. Here, $q$ is the
particle's electric charge, $m$ being its mass, $E$ the energy. The
forward-in-time condition $\dot{t}>0$ entails%
\begin{equation}
X>0.  \label{ft}
\end{equation}

The electric Coulomb potential%
\begin{equation}
\varphi =\frac{Q}{r}\text{.}  \label{pot}
\end{equation}%
Hereafter, we use notations $\varepsilon =\frac{E}{m},$ $\tilde{q}=\frac{q}{m%
}$. Then,%
\begin{equation}
U(r)=(\varepsilon -\frac{\tilde{q}Q}{r})^{2}-(1-\frac{2M}{r}+\frac{Q^{2}}{%
r^{2}})\text{.}  \label{U}
\end{equation}

Then, the possible turning points $r_{t}$ can be found from the condition $%
P=0$, whence

\begin{equation}
r_{t}^{\pm }=\frac{1}{\varepsilon ^{2}-1}(\varepsilon \tilde{q}Q-M\pm \sqrt{C%
})\text{,}  \label{tp}
\end{equation}%
\begin{equation}
C=(M-\varepsilon \tilde{q}Q)^{2}+(1-\tilde{q}^{2})Q^{2}(\varepsilon ^{2}-1).
\label{C}
\end{equation}

\section{Scenario of decay}

In what follows, we will be mainly interested in the situation when decay
occurs near the singularity, so the point of decay $r_{0}\rightarrow 0$.
Then, for fixed $q_{0}$ and $E_{0}$, the forward-in-time condition (\ref{ft}%
) requires $q_{0}<0$. Assuming also that particle moves from infinity, so $%
\varepsilon _{0}>1$, we see from (\ref{tp}) that no more than one turning
point $r_{t}^{+}$ can exist, provided $\left\vert \tilde{q}_{0}\right\vert
<1 $.

Let particle 0 with $\varepsilon >1$ fall from infinity. In some point $%
r_{0} $ it decays to two new fragments 1 and 2. We assume the conservation
of the energy and electric charge in the point of decay, so%
\begin{equation}
E_{0}=E_{1}+E_{2}\text{,}  \label{e}
\end{equation}%
\begin{equation}
q_{0}=q_{1}+q_{2}\text{.}  \label{q}
\end{equation}%
The necessary condition that makes decay possible is 
\begin{equation}
m_{0}\geq m_{1}+m_{2}\text{.}  \label{m}
\end{equation}

For given characteristics $E_{0},m_{0}$, $q_{0}$ of particle 0, one can
solve (\ref{e}), (\ref{q}) with (\ref{P}) taken into account. We can take
advantage of already obtained results \ - see eqs. (19) -\ (25) of Ref. \cite%
{centr}. Only minimum changes are required: (i) instead of indices 3, 4 we
use here 1,2, (ii) the quantity $X$ is defined according to (\ref{P})
instead of eq. (5) of \cite{centr}.

Then,%
\begin{equation}
X_{1}=\frac{1}{2m_{0}^{2}}\left( X_{0}\Delta _{+}+P_{0}\delta \sqrt{D}%
\right) ,  \label{X1}
\end{equation}%
\begin{equation}
X_{2}=\frac{1}{2m_{0}^{2}}\left( X_{0}\Delta _{-}-P_{0}\delta \sqrt{D}%
\right) ,  \label{X2}
\end{equation}%
\begin{equation}
P_{1}=\left\vert \frac{P_{0}\Delta _{+}+\delta X_{0}\sqrt{D}}{2m_{0}^{2}}%
\right\vert \text{,}
\end{equation}%
\begin{equation}
P_{2}=\left\vert \frac{P_{0}\Delta _{-}-\delta X_{0}\sqrt{D}}{2m_{0}^{2}}%
\right\vert ,  \label{P2}
\end{equation}%
where $\delta =\pm 1$,%
\begin{equation}
\Delta _{\pm }=m_{0}^{2}\pm (m_{1}^{2}-m_{2}^{2}),
\end{equation}%
\begin{equation}
D=\Delta _{+}^{2}-4m_{0}^{2}m_{1}^{2}=\Delta _{-}^{2}-4m_{0}^{2}m_{2}^{2}.
\label{D}
\end{equation}%
It follows from (\ref{m}) that $D\geq 0$. The equality holds if $%
m_{0}=m_{1}+m_{2}$ only.

The solutions (\ref{X1}) - (\ref{D}) are classified according to 4
quantities $(\sigma _{2}$, $h_{2}$, $h_{1}$, $\delta )$. The corresponding
allowed combinations are listed in eq. (30) of \cite{centr}. (A reader
should bear in mind that the role of particle 3 in \cite{centr} is played
now by particle 2). Here, 
\begin{equation}
h_{1}=sgnH_{1}\text{, }H_{1}=\Delta _{+}\sqrt{f}-2m_{1}X_{0}\text{,}
\end{equation}%
\begin{equation}
h_{2}=sgnH_{2}\text{, }H_{2}=\Delta _{-}\sqrt{f}-2m_{2}X_{0}\text{.}
\label{H2}
\end{equation}

We are interested in the situation when particle 2 escapes. If it moves
after decay immediately to infinity, $\sigma _{2}=+1$. We will consider this
type of scenario first. (Afterwards, we will also discuss an alternative
scenario when particle 2 bounces back from the potential barrier.)

Then, according to \cite{centr}, there are only two possibilities: $%
1(+,+,+,-)$ and $2(+,+,-,-)$. Thus $\delta =-1$ and we should also have 
\begin{equation}
H_{2}>0  \label{h2}
\end{equation}%
while $H_{1}$ can have any sign.

\section{Energy extraction}

The Penrose process implies that particle 1 moves toward the center with $%
E_{1}<0$, whereas particle 2 escapes to infinity with $E_{2}>0$. In doing
so, $\sigma _{1}=-1$ and $\sigma _{2}=+1$. Our goal is to obtain the energy
extraction as large as possible. According to (\ref{X}),%
\begin{equation}
E_{2}=X_{2}+\frac{q_{2}Q}{r_{0}}.  \label{e2}
\end{equation}

Taking into account (\ref{ft}), we see that if $q_{2}>0$ and $%
r_{0}\rightarrow 0$, the energy $E_{2}$ is unbounded. Is it possible to
achieve this goal in our scenario?

When $r_{0}\rightarrow 0,$ condition (\ref{ft}) requires $q_{0}<0$. Thus we
should have%
\begin{equation}
q_{0}<0\text{, }q_{2}>0\text{.}
\end{equation}%
Further, we should consider two different cases depending on whether or not
there is a turning point for particle 0.

\subsection{No turning point}

This case is realized if $\left\vert \tilde{q}_{0}\right\vert >1$ since both
roots (\ref{tp}) become negative. Then, we can take the limit $%
r_{0}\rightarrow 0$ directly. In this limit, it follows from (\ref{X}), (\ref%
{P}), (\ref{pot}) that%
\begin{equation}
X_{0}\approx \frac{\left\vert q_{0}\right\vert Q}{r_{0}},  \label{xoa}
\end{equation}%
\begin{equation}
P_{0}\approx \frac{Q}{r_{0}}\sqrt{q_{0}^{2}-m_{0}^{2}}.  \label{pa}
\end{equation}%
Now, using (\ref{X2}), (\ref{e2}, )(\ref{xoa}) and (\ref{pa}), one obtains
in the main approximation%
\begin{equation}
E_{2}\approx \frac{Q}{r_{0}}[q_{2}+\frac{1}{2m_{0}^{2}}(\left\vert
q_{0}\right\vert \Delta _{-}+\sqrt{q_{0}^{2}-m_{0}^{2}}\sqrt{D})]  \label{no}
\end{equation}%
and $E_{2}\rightarrow \infty $ when $r_{0}\rightarrow 0$.

Now,%
\begin{equation}
H_{2}\approx \frac{Q}{r_{0}}(\Delta _{-}-2m_{2}\left\vert q_{0}\right\vert ).
\end{equation}%
Eq. (\ref{h2}) is valid, if%
\begin{equation}
\left\vert q_{0}\right\vert <\frac{\Delta _{-}}{2m_{2}}\text{.}
\end{equation}%
It is consistent with $\left\vert \tilde{q}_{0}\right\vert >1$.

This is not the end of story. We must check that particle 2 does escape, so
it does not have a new turning point for all $r>r_{0}$, $U>0$. We should
verify that $U_{2}(r)>U_{2}(r_{0})$ for any $r>r_{0}$. Using (\ref{U}), it
is easy to find that%
\begin{equation}
U_{2}(r)-U_{2}(r_{0})\approx \frac{(r-r_{0})}{rr_{0}^{2}}B\text{, }
\end{equation}%
\begin{equation}
B=Q^{2}[1+\tilde{q}_{2}^{2}+\frac{r_{0}(1-\tilde{q}_{2}^{2})}{r}+\frac{q_{2}%
}{m_{2}^{2}m_{0}^{2}}(\left\vert q_{0}\right\vert \Delta _{-}+\sqrt{%
q_{0}^{2}-m_{0}^{2}}\sqrt{D})]-2Mr_{0}.
\end{equation}%
Obviously, if $r_{0}\rightarrow 0$, $B>0$ for any $r>r_{0}$. Thus, 
\begin{equation}
U_{2}(r)>U_{2}(r_{0})>0  \label{U2}
\end{equation}%
and there are no additional turning points, so particle 2 escapes to
infinity freely.

\subsection{Decay in the turning point}

Let us suppose now that the turning point for particle 0 does exist. As we
try to obtain the maximum possible $E_{2}$, it makes sense to choose (for
given values of other parameters) the minimum possible value of $r_{0}$. To
this end, we put $r_{0}=r_{t}^{(+)}$. We assumed (as is explained above)
that $q_{0}<0$. Then, according to eqs. (\ref{tp}), (\ref{C}) the existence
of a turning point requires $\left\vert \tilde{q}_{0}\right\vert \leq 1$. In
this point we have $P_{0}=0$ by definition, so it follows from (\ref{X1}) - (%
\ref{P2}) that%
\begin{equation}
X_{1}=\frac{X_{0}\Delta _{+}}{2m_{0}^{2}}\text{,}  \label{x1}
\end{equation}%
\begin{equation}
X_{2}=\frac{X_{0}\Delta _{-}}{2m_{0}^{2}},  \label{x2}
\end{equation}%
\begin{equation}
P_{1}=P_{2}=\frac{X_{0}}{2m_{0}^{2}}\sqrt{D}\text{,}  \label{p2}
\end{equation}%
where now%
\begin{equation}
X_{0}=E_{0}+\frac{\left\vert q_{0}\right\vert Q}{r_{0}},  \label{x0}
\end{equation}%
\begin{equation}
E_{2}=\frac{E_{0}\Delta _{-}}{2m_{0}^{2}}+\frac{Q}{r_{0}}(q_{2}+\left\vert
q_{0}\right\vert \frac{\Delta _{-}}{2m_{0}^{2}})\text{.}  \label{e2t}
\end{equation}%
In doing so, $X_{0}>0$ for any $r_{0}$ due to $q_{0}<0$, so condition (\ref%
{ft}) holds. As we want to minimize $r_{0}$, we choose%
\begin{equation}
\left\vert \tilde{q}_{0}\right\vert =1-\beta \text{, }\beta \ll 1\text{.}
\label{qb}
\end{equation}%
Then, it follows from (\ref{tp}) that 
\begin{equation}
r_{0}\approx \frac{Q^{2}\beta }{M+\varepsilon Q}\text{.}
\end{equation}

can be made as small as one likes. As a result, we have from (\ref{e2t}) that%
\begin{equation}
E_{2}\approx \frac{(M+\varepsilon Q)}{Q\beta }(q_{2}+\left\vert
q_{0}\right\vert \frac{\Delta _{-}}{2m_{0}^{2}}).
\end{equation}%
When $\beta \rightarrow 0$, $E_{2}\rightarrow \infty $, so the SPP does
exist.

Eq. (\ref{U2}) is valid in the case under consideration as well. It is worth
noting that if $q=0,$ the expression (\ref{tp}) coincides with eq. (13) of 
\cite{naked}. However, we saw that in both versions of the scenario under
consideration (with a turning point or without it) it is essential for the
SPP that $q_{0}\neq 0$. Thus, this process is possible for charged particles
and is absent for neutral ones.

\section{Alternative type of scenario}

For completeness, we must consider the case when \ particle 2 after decay
moves in the same direction as particle 1, so $r$ continues to decrease.
However, immediately after decay particle 2 bounces back from the potential
barrier. This means that $r_{0}$ is the turning point for particle 2, so $%
P_{2}=0$ and, therefore,%
\begin{equation}
X_{2}=m_{2}\sqrt{f}\text{.}
\end{equation}

When $r_{0}\rightarrow 0$, 
\begin{equation}
X_{2}\approx \frac{m_{2}Q}{r_{0}}\text{.}  \label{x2a}
\end{equation}

According to (\ref{P2}), we must take $\delta =+1$ and%
\begin{equation}
P_{0}\Delta _{-}=X_{0}\sqrt{D}\text{.}
\end{equation}%
It is easy to check that this is equivalent to $H_{2}=0$ in (\ref{H2}), so%
\begin{equation}
\Delta _{-}\sqrt{f}=2m_{2}X_{0}\text{.}  \label{h20}
\end{equation}

If $q_{0}<0$, there is no turning point for particle 0. In the limit $%
r_{0}\rightarrow 0$, $X_{0}\approx \frac{\left\vert q_{0}\right\vert Q}{r_{0}%
}$, $\sqrt{f}\approx \frac{Q}{r_{0}}$ and we obtain from (\ref{h20}) 
\begin{equation}
\left\vert q_{0}\right\vert \approx \frac{\Delta _{-}}{2m_{2}}\text{.}
\end{equation}

If $q_{0}>0$, choosing $r_{0}$ to be a turning point for particle 0 as well,
we have $X_{0}=m_{0}\sqrt{f}$, so it follows from (\ref{h20}) that $\Delta
_{-}=2m_{0}m_{2}$, whence $m_{0}=m_{1}+m_{2}$. In both cases, according to (%
\ref{x2a}),%
\begin{equation}
E_{2}\approx \frac{Q(m_{2}+q_{2})}{r_{0}}\text{.}
\end{equation}

Here, we should take $q_{2}>-m_{2}$. Thus the SPP does exist in this case
also.

\section{Flat space-time limit}

Decay in the case of the flat space-time is of special interest. More
precisely, we put $M=Q=0$ in the metric thus neglecting the influence of the
electromagnetic field on space-time. However, we take into account the
electric charge in equations of motion. Again, let particle 0 decay in the
point $r_{0}$. Then, eqs. (\ref{X1}), (\ref{X2}) are now valid with $f=1$.
Now we will show that the SPP is still possible for a finite value of $%
\left\vert q_{0}\right\vert $ (the corresponding scenarios were overlooked
in our previous paper \cite{flat}).

If the turning point $P_{0}=0$ exists, its coordinate is given by 
\begin{equation}
r_{0}=\frac{q_{0}Q}{E_{0}-m_{0}}\text{.}  \label{r0f}
\end{equation}

Here, it is assumed that $E_{0}>m_{0}$. Now there are two different cases.

\subsection{No turning point}

Let $q_{0}<0$. Then, the turning point for particle 0 is absent. We want
particle 2 to escape to infinity, so the turning point for particle 2 should
be absent as well, \thinspace $q_{2}=-\left\vert q_{2}\right\vert <0$.
Proceeding along the same lines as before, we obtain for $r_{0}\rightarrow 0$%
\begin{equation}
E_{2}\approx \frac{Q}{r_{0}}[\frac{\left\vert q_{0}\right\vert }{2m_{0}^{2}}%
(\Delta _{-}+\sqrt{D})-\left\vert q_{2}\right\vert ]  \label{e2f}
\end{equation}

This expression differs from (\ref{no}) since in (\ref{no}) it was implied
that $f\sim \frac{Q^{2}}{r_{0}^{2}}\rightarrow \infty $, whereas now $f=1.$

The positivity of $E_{2}$ for small $r_{0}$ requires%
\begin{equation}
\left\vert q_{2}\right\vert <\frac{\left\vert q_{0}\right\vert }{2m_{0}^{2}}%
(\Delta _{-}+\sqrt{D})\text{.}
\end{equation}

Formally, (\ref{e2f}) grows indefinitely when $r_{0}\rightarrow 0$.
Actually, as we neglected the corresponding term $\frac{Q^{2}}{r^{2}}$ in
the RN metric, there is an additional constraint on $r_{0}$. Restoring
dimensionality, we have $r_{0}\gg \frac{\sqrt{G}Q}{c^{2}}$, so there is
restriction $E_{2}\ll \frac{c^{2}}{\sqrt{G}}[\frac{\left\vert
q_{0}\right\vert }{2m_{0}^{2}}(\Delta _{-}+\sqrt{D})-\left\vert
q_{2}\right\vert ]$ that is rather weak in the case $G\rightarrow 0$. Now, 
\begin{equation}
H_{2}=\Delta _{-}-2m_{2}(E_{0}+\frac{\left\vert q_{0}\right\vert Q}{r_{0}}).
\label{H2+}
\end{equation}%
It is seen that for a fixed $\left\vert q_{0}\right\vert $ and $%
r_{0}\rightarrow 0$ eq. (\ref{h2}) cannot be fulfilled. It means that the
scenario under discussion cannot be realized, \ so particle 2 falls in the
center along with particle 1 and does not escape. The situation changes if
we take $\left\vert q_{0}\right\vert =\alpha r_{0}$, where $\alpha =O(1)$.
Then, we have from (\ref{H2+}) that%
\begin{equation}
\alpha <Q^{-1}(\frac{\Delta _{-}}{2m_{2}}-E_{0}).  \label{al}
\end{equation}

\subsection{Decay in the turning point}

Now, $P_{0}=0$, so eqs. (\ref{x1}) - (\ref{x0}) apply. As \ a particle falls
from infinity, $E_{0}>m_{0}$. According to (\ref{r0f}), we must also take $%
q_{0}>0$. For particle 2 we take $q_{2}<0$ to exclude a possible turning
point after decay. As we want $r_{0}\rightarrow 0$, it is seen from (\ref%
{r0f}) that we must assume $q_{0}\rightarrow 0$. Then, in eq. (\ref{e2t})
the main contribution becomes negative, so the SPP is absent in this case.

\subsection{Alternative scenario}

Is the alternative scenario possible in the flat case? It implies that
particle 2 bounces back from the turning point. Then, $P_{2}=0$, so%
\begin{equation}
X_{2}=m_{2}\text{.}  \label{xm}
\end{equation}%
Now, one should put $f=1$ in eq. (\ref{h20}) typical of a turning point.
Then, putting there $X_{0}\approx \frac{\left\vert q_{0}\right\vert Q}{r_{0}}
$, we obtain%
\begin{equation}
\left\vert q_{0}\right\vert \approx \frac{r_{0}}{Q}\frac{\Delta _{-}}{2m_{2}}%
.
\end{equation}%
Thus for a fixed $q_{0}$ we have%
\begin{equation}
E_{2}=m_{2}+\frac{q_{2}Q}{r_{0}}\text{,}
\end{equation}%
where we took into account (\ref{xm}). Then,%
\begin{equation}
E_{2}\approx \frac{q_{2}\Delta _{-}}{2m_{2}\left\vert q_{0}\right\vert Q}%
\text{,}
\end{equation}%
where it is assumed $q_{2}>0$. If $q_{0}\rightarrow 0$, $E_{2}\rightarrow
\infty $, so the SPP occurs. As now $X_{2}$ is monotonically increasing
function of $r$, there are no other turning points and particle 2 escapes$.$

\section{Discussion: comparison of different types of high energy processes}

It is instructive to compare the obtained results with those already
existing in literature, both for the rotating and static charged case.
Usually, the outcome of particle collisions depends strongly on several
factors: (i) the parameters of geometry, (ii)\ parameters of particles,
(iii) direction of motion. Let us remind a reader at first the situation for
processes with neutral particles in the background of rotating black holes.
If one wants to have unbounded $E_{c.m.}$ for collision of particles falling
from infinity, a black hole, typically, has to be extremal. This implies a
special relation between the mass $M$ and angular momentum $a$ of a black
hole: $a=M$ (in corresponding units) for the Kerr metric. Also, one of
particles should be fine-tuned (so-called critical) which implies a special
relation between the parameters of a particle itself \cite{ban}. However,
even with unbounded $E_{c.m.}$, the energy $E$ remains quite modest \cite{j}
- \cite{z}. It can be significantly increased if a fine-tuned particle is
outgoing \cite{sch}, \cite{maxpiran}. However, the energy $E$ remains
bounded anyway.

If collision occurs in the field of a naked singularity, there is no need in
fine-tuning particle parameters but the geometry itself should be on the
threshold of forming a horizon \cite{inf}, \cite{tzw}, \cite{n}. For
example, for the Kerr metric $a$ has to be only slightly bigger than $M$.
Alternatively, one can try to relax the requirement of having the extremal
black hole and consider the Schnittman scenario \cite{sch} (collision
between an infalling and outgoing particles) even for the nonextremal one.
However, in this case, the input of energy should be very large from the
very beginning, that depreciates the value of the process \cite{pir15}.
Meanwhile, in the current setup, neither geometry nor particles should be
fine-tuned. Also, the relative direction of motion is irrelevant. Although
details of the process look different for each separate scenario, the SPP
turns out to be possible almost for all scenarios considered above.

One can also compare our problem with collisions of charged particles near
the RN black hole. One should require one of infalling particles to be a
critical as well. If both particles are usual (not fine-tuned), neither
unbounded $E_{c.m.}$ or unbounded $E$ are possible \cite{rn}, \cite{nem}, 
\cite{hlz}. And, for the critical particle there is a bound on the electric
charge $q<Ze$, where $e$ is the value of an elementary charge and $Z=170$
comes from quantum electrodynamics. As the energy of a particle produced in
collision turns out to be proportional to the charge $q$ \cite{rn}, this
gives rise to the restriction of the efficiency of the PP. But this
circumstance is irrelevant in the case under discussion since high energies
can be obtained even for a modest electric charge.

And, the last but not the least, there is no necessity in arranging the
collisional Penrose process for achieving the SPP. A standard decay of one
particle is quite sufficient. Thus, a naked singularity provides us with the
most favorable situation for gaining energy and accelerating particles to
ultrahigh energies.

\section{Conclusions}

As is known, the electric charge of astrophysical objects is rather small.
(Although it can, in principle, lead to observable effects \cite{tur}.)
However, in some aspects the electrical charge can model what happens in
more complicated realistic astrophysical systems with rotation. Meanwhile,
the RN metric is much easier for analysis than, say, the Kerr metric
describing the vacuum solution of the Einstein equations with rotation. In
this sense, the results obtained in the present paper can be of some use for
the analysis of more realistic processes.

They are also interesting on their own right. We found one more type of
systems for which the SPP is possible. It exists in almost all scenarios
considered above. In the configuration considered in the present paper (i)
the SPP can occur without fine-tuning at all and (ii) to gain large $E$,
there is no need to have large $q$. More precisely, fine-tuning is required
for the subcase with the turning point (\ref{qb}) only but is absent if such
a point is missing. One more difference between the present version of SPP
and that near a black hole consists in that now the original particle needs
not be ultrarelativistic having any finite value $\varepsilon >1$ (cf. Sec.
IV C3 in \cite{axis}). In this sense, the present version of the SPP is less
restrictive than the previous one typical of the BSW effect. We would also
like to stress that the effect under discussion concerns indefinitely large
Killing energy $E$, whereas $E_{c.m.}$ is irrelevant since we considered
particle decay, not collision.

In combination with the previous results \cite{ruf}, \cite{den}, \cite{rn}, 
\cite{df}, \cite{flat}, this means that any type of the Reissner-Nordstr\"{o}%
m space-time \ (black hole, flat space-time as its limit, naked singularity)
is pertinent to the SPP.

\end{document}